# Phononically shielded multi-wavelength photonic-crystal membrane for cavity quantum optomechanics

HANBING LI,[1,2] DOUDOU WANG,[1,2] QUANSEN WANG,[1,2] QIANG ZHANG,[1,2] AND YONGMIN LI[1,2,*]

[1]State Key Laboratory of Quantum Optics and Quantum Optics Devices, Institute of Opto-Electronics, Shanxi University, Taiyuan 030006, China

[2]Collaborative Innovation Center of Extreme Optics, Shanxi University, Taiyuan, Shanxi 030006, China

[*]yongmin@sxu.edu.cn

**Abstract:** We propose and design a stoichiometric silicon-nitride membrane resonator featuring highly reflective at multi-wavelengths and high mechanical quality factor. The membrane resonator has a thickness of 100 nm and 2D-photonic and phononic crystal patterns. By designing concentric holes of suitable radius on both sides of the membrane, high reflectivity at multi-wavelengths can be achieved. In particularly, the simulation shows that high reflectivity can be realized at telecommunications wavelength and alkaline atoms absorption lines, with reflectivity of 99.76% at 852 nm, and 99.98% at 1054 nm, and 99.96% at 1566 nm, respectively. The designed device can find useful applications in cavity optomechanical system to realize quantum frequency conversion and precise quantum measurement, and other field of quantum information processing tasks.

## 1. Introduction

Cavity optomechanics exploits the radiation pressure interaction between the optical field and the mechanical resonators [1,2]. It has wide applications from fundamental quantum physics to quantum measurements, and quantum information processing [3]. By circulating optical fields in cavities, the interaction between the light and mechanical resonators can be greatly enhanced. The confining of the optical fields in cavities also enables a number of interesting physical phenomena ranging from the motional ground state cooling of the resonator, to optomechanically induced transparency, squeezing and entanglement of light fields, and mechanical oscillators and transversion between microwave and optical light, and quantum sensing, etc. The dielectric membrane resonators based cavity optomechanics system where the optical and mechanical resonators can be relatively independently designed and optimized, have attracted a lot of interests [4].

Silicon nitride (SiN) membranes, features low optical absorption and high mechanical quality factors have been widely used in many membrane-in-the-middle cavity optomechanical system. By designing photonic crystals structure [5–8]on the membrane, high reflectivity can be achieved, and. In this way, the membrane can be acted as both the resonator and one of the end mirrors of the optomechanical system. In contrast to the membrane-in-the-middle system, such a configuration can enable a compact optomechanical system and enhance the optomechanical coupling rate $g_0$.

Recently, several experiments have explored the photonic crystal membranes in cavity optomechanics [9–14]. Moura et al. [15] presented centimeter-scale suspended photonic crystal mirrors and obtained reflectivity higher than 90 % at 1550 nm. Enzian et al. [16] presented a stoichiometric silicon-nitride membrane, which combined both the photonic and phononic crystal structures and has reflectivity of 99.89% and mechanical quality factor of $2.9 \times 10^7$ at room temperature. Agrawal et al. [17] realized a 99% reflectivity focusing metamirror using non-periodic photonic crystal patterning in a $Si_3N_4$ membrane. Very recently, Zhou et al. [18] reported ultrahigh reflectivity photonic crystal membranes, which can reach reflectivity up to 99.982% using a hexagonal lattice. They find that the exagonal lattices have a wider spectral reflection range and reflectance, less susceptibility to manufacturing imperfections in comparison to the square lattices.

At present, most of the research of the photonic crystal membranes focus on the realization of high reflectance at a single wavelength [15–18]. Extending the high reflectance ability at a single wavelength to multiple wavelength range is useful to enrich the toolbox of the cavity optomechanical systems. For instance, a two-cavity optomechanics system can be constructed based on two-wavelength photonic crystal membranes to achieve quantum frequency conversion between quantum memory wavelength at 800 nm band and telecom wavelength at 1.5 micron.

In this work, we propose and design a stoichiometric silicon-nitride membrane resonator featuring highly reflective at multi-wavelengths and high mechanical quality factor. The membrane resonator has a thickness of 100 nm and 2D-photonic and phononic crystal patterns. By designing concentric holes of suitable radius on both sides of the membrane, high reflectivity at multi-wavelengths is achieved. In particularly, the simulation results shows that high reflectivity can be realized at telecommunications wavelength and alkaline atoms absorption lines, with reflectivity of 99.76% at 852 nm, and 99.98% at 1054 nm, and 99.96% at 1566 nm, respectively. The designed devices can be exploited to construct a compact optomechanical system with high optomechanical coupling rate, which can find potential applications in cavity optomechanical system to realize quantum frequency conversion, preparation of nonclassical optical and mechanical states, as well as ultrasensitive quantum measurement.

The high mechanical quality factor is enabled by dissipation dilution and soft clamping [19]. The increased reflectivity enables enhanced optomechanical coupling rates and cooperativities to be reached in membrane-in-the-middle or Fabry-Perot setups. Such enhanced operativity could enable improved squeezing and sensing capabilities, as well as ground state cooling starting from room temperature.

This article is organized as follows. In section 2, we present the design and simulation results of the membrane resonator in details. In section 3, we present the potential application of the designed phononically shielded multi-wavelength photonic-crystal membrane on cavity quantum optomechanics. Finally, we give a summary.

## 2. Phononically shielded multi-wavelength photonic-crystal Si$_3$N$_4$ membrane resonator

### 2.1. Design and parameters

The basic structures of our Si$_3$N$_4$ membrane resonator consist of both the photonic crystal and the phononic crystal patterns [16,20]. Based on this structure, we design and optimize the phononic crystal and photonic crystal structures to enable the membrane resonator possessing both high mechanical quality factors and high reflectivity at multi-wavelength.

As shown in Figure 1, the squared Si$_3$N$_4$ membrane has a thickness of 100 nm and side length of 3.8 mm, and it is suspended on top of a 10×10 mm Si substrate with the thickness of 500 μm. The phonon crystal consists of a honeycomb lattice of circular holes with lattice constant 173.2 μm. The radius of the dominant circular holes is 45 μm, and the radius of the secondary holes that surrounding the photonic crystal is 23.85 μm. A circular geometric defect with a radius of 95 μm is introduced in the center of the phonon crystal structure to localize the vibration of the mechanical mode in the bandgap of phonons crystal. The defect contains the photonic crystal structure that is made of a hexagonal lattice of concentric horn like holes with lattice constant of 1695 nm. The radius of the concentric holes are 33 nm and 622.2 nm, respectively, and the depth of the holes is 30 nm and 70nm respectively, as shown in Fig. 2(a).

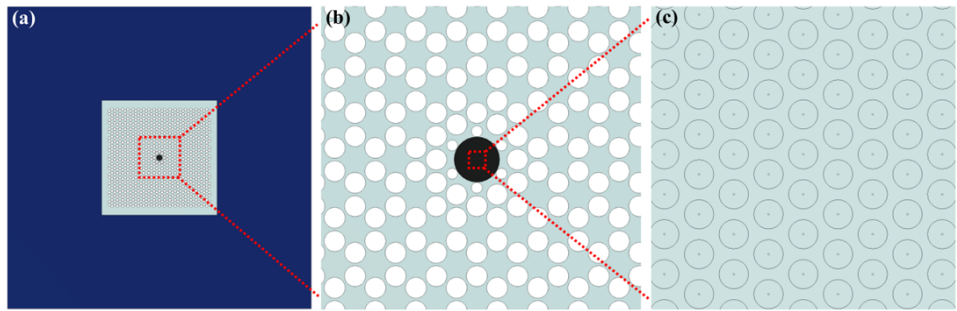

Fig. 1. Schematic diagram of the phononically shielded multi-wavelength photonic-crystal Si$_3$N$_4$ membrane resonator. (a) The blue region denotes the silicon frame, and the light grey-green region represents the Si$_3$N$_4$ membrane, which contains two kinds of structures, the centered hexagonal photonic crystal structure (black) surrounded by the honeycomb phononic crystal structure (light grey-green). (b) Zoom of the structure of the phononic crystal. (c) Zoom of the photonic crystal structure.

*2.2 Simulation*

Numerical simulations of both the optical and the mechanical characteristics of the membrane resonator are performed using finite element method (FEM) with Comsol Multiphysics [21].

2.2.1 Optical characteristics of the photonic crystal structures

We simulate the reflectance of multi-wavelength photonic crystal structures by using the physical field of electromagnetic wave frequency domain (EWFD) of the COMSOL. A single unit-cell with periodic boundaries that modelling an infinite array and a plane-wave inputting to the photonic crystal structures are employed in the simulation model. The simulation uses the parameters mentioned above and the index of refraction of $Si_3N_4$: $Re(n) = 1.98$ and $Im(n) = 5 \times 10^{-6}$ [22]. Moreover, the port boundary conditions for transmitting the incident waves, absorbing the transmitted and reflected waves are set [20].

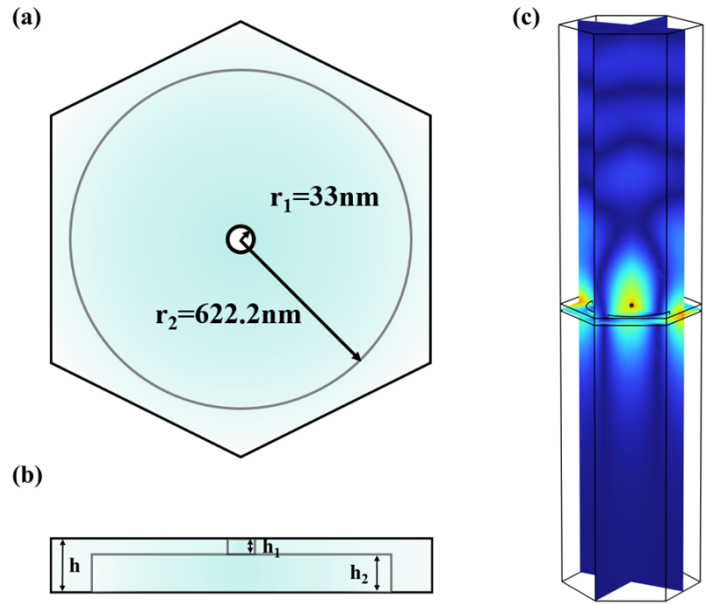

Fig. 2. (a) Three-dimensional top view of a single unit cell of the photonic crystal. The radius of the small hole for the photonic crystals is $r_1 = 33$ nm and that of the large hole is $r_2 = 622.2$ nm. (b) Front view of single unit cell. The thickness of the cell is $h = 100$ nm, and the two holes have the different thickness of $h_1 = 30$ nm, $h_2 = 70$ nm. (c) The electric field distribution diagram using the COMSOl plane wave simulation.

Figure 3 shows the reflectivity of the photonic crystal structure of the membrane as a function of the wavelength simulated by the FEM. Here, the side of the photonic crystal structures with a smaller hole radius is set as the entrance port. From the figure, we can find that the photonic crystal structure of the membrane exhibits near-unity high reflectivity at multiple wavelengths. The reflectivity is 99.76% at 852 nm, 99.98% at 1054 nm, and 99.96% at 1566 nm, respectively. The wavelength of 852 nm and 1566 nm falls in the telecom windows. The wavelength of 1054 nm falls in the waveband of free-space optical communication. Note that the wavelength of 852 nm is also the D2 line of Cs atoms, which is suitable for quantum memory and long distance quantum communication.

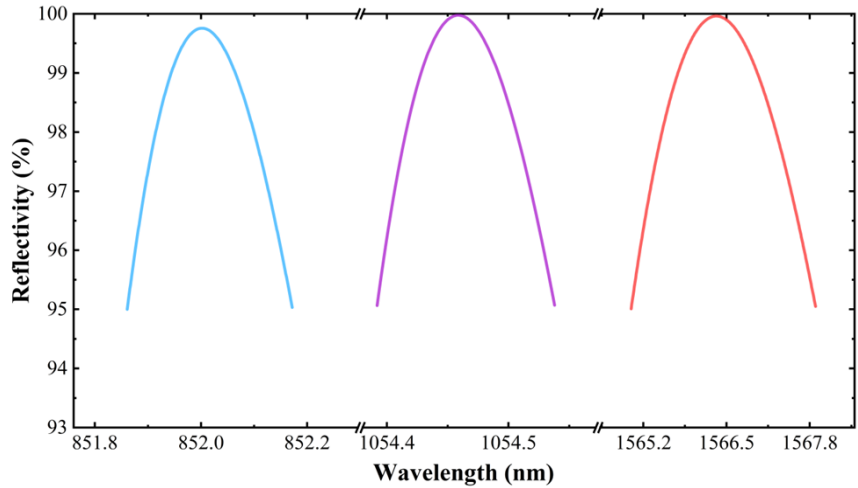

Fig. 3. The reflectivity and transmission of the photonic crystal structure of the SiN membrane as a function of the wavelength simulated by the FEM. A plane wave at normal incidence and infinite extension of the photonic crystal structure is adopted. The reflectivity reaches 99.76% at 852 nm, 99.98% at 1054.5 nm, and 99.96% at 1566.3 nm, respectively.

Because the photonic crystal structure is up and down asymmetric. The characteristics of the reflectivity (transmission) is not identical when the light field impinges on the photonic crystal structures membrane from different sides. Figure 3 shows the reflectivity and transmission when the plane wave input on the membrane from upper side (Fig. 2(b)). If the plane wave incidents on the membrane from the below side, the reflectivity is 40.49% at 852 nm, and 99.98% at 1054.5 nm, and 99.97% at 1566.3 nm, respectively. For some applications, the light fields will incident on the membrane from both sides (please refers to Sec.3), in this case, we can select the appropriate planes of the incidence to ensure a high reflectivity. For instance, the light at 852 nm can incident on the membrane from the upper side, whereas the light at 1566 nm can be impinged from the other side.

2.2.2 Mechanical properties of the membrane resonator

We use the stationary and the eigenfrequency to study mechanical properties of the membrane resonator in COMSOL. The eigenfrequency of the membrane resonator is simulated by the membrane physical field in COMSOL, and the band gap of the phonon crystal is simulated by the solid mechanics physical field [23,24]. The whole structure of membrane is used to simulate the eigenfrequency. Similar to the reflectivity of the photonic crystal, a single unit-cell with periodic boundaries that modelling an infinite array are employed to simulate the band gap diagram of the phonon crystal. For the physical field of the membrane and the solid mechanics, we choose a three-dimensional model design. By using the parameters of the phonon crystal structures mentioned above and SiN: the Young's modulus of 270 GPa, Poisson's ratio of 0.27, and density of 3200 kg/m$^3$, we get the physical properties of the SiN membrane In addition, the simulation of the eigenfrequency requires fixed constraint boundary conditions, and bandgap graph simulation requires both the fixed constraint boundary and periodic boundary conditions [20].

Figure 4(a) depicts the simulated mode shape of the mechanical defect mode we studied, which is the lowest-order mechanical defect mode with a vibration frequency of $\omega_m = 1.34$ MHz. Figure 4(b) depicts the simulated band gap diagram of the phonon crystal structure by parametric sweep and two band gaps (light blue part) are visible. The first band gap is relatively broad that ranging from 1.168 to 1.406 MHz, and the second band gap ranging from 2.287 to 2.314 MHz is narrower. It is clear that the vibration frequency of the lowest-order mechanical defect mode resides in the first band gap. Therefore, this mechanical mode will be immune to the external mechanical vibration and a high mechanical quality factor can be achieved.

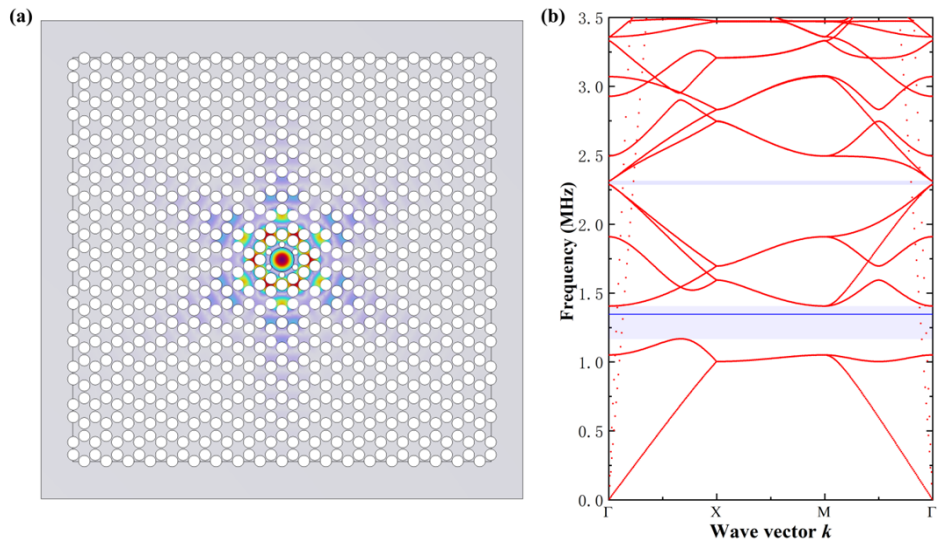

Fig. 4. Mechanical properties of the phononically shielded membrane resonators. (a) Mode shape diagram of the mechanical defect mode, which shows the lowest-order mechanical defect mode has a frequency of 1.34 MHz. (b) Band gap diagram of the phonon crystal structure. The light blue part corresponds to the band gap of the phonon crystal, and the blue line represent the vibration frequency of lowest-order mechanical defect mode ($\omega_m = 1.34$ MHz), which lies in the middle of the phonon gap band.

The mechanical quality factor characterizes the mechanical loss of resonator and is defined as the ratio of the energy stored in the mechanical resonator to the energy lost per

unit vibration period [25,26]

$$Q_m = \frac{E_{stored}}{-\frac{dE_{stored}}{dt} \cdot T}, \quad (1)$$

where $E_{stored}=A_0^2 e^{-2tIm(\omega_m)}$ is the mechanical energy stored in the mechanical resonator, and $T = \frac{2\pi}{Re(\omega_m)}$ is the vibration period. By substituting these two expressions into Eq. (1), we can get

$$Q_m = \frac{Re(\omega_m)}{2 \cdot Im(\omega_m)}, \quad (2)$$

where $Re(\omega_m)$ and $Im(\omega_m)$ is the real and imaginary parts of the eigenfrequency, respectively. From Eq. (2), the mechanical quality factor $Q_m$ is estimated to be $1.59 \times 10^8$.

In addition, we can estimate the effective mass $m_{eff}$ of the designed resonator via the mode shape of the membrane resonator [16]

$$m_{\text{eff}} = \iint_S dxdy \rho(x,y) d \left| \frac{D(x,y)}{D(x_0,y_0)} \right|^2, \quad (3)$$

where $\rho(x,y)$ is the density of membrane, $d$ is the thickness of membrane resonator, $z(x,y)$ is the out-of-plane displacement of the mechanical mode, $D(x,y)/D(x_0,y_0)$ denotes the normalized displacement field of the mechanical mode that can be obtained via the mode shape of the membrane resonator [16]. Due to the introduction of photonic crystal structure, the effective mass of the membrane oscillator is changed from 9.48 ng to 15.17 ng.

### 2.2.3 Optomechanical cooperativity

High optomechanical cooperativity is crucial to the optomechanical systems and their applications [27]. The optomechanical cooperativity $C$ in cavity optomechanical systems is defined as

$$C \equiv 4g_0^2 n_c / \kappa \Gamma_m, \quad (4)$$

where

$$g_0 = G x_{\text{zpf}}, \quad x_{\text{zpf}} = \sqrt{\frac{\hbar}{2m_{\text{eff}}\Omega_m}}, \quad (5)$$

where $g_0$ is the single-photon optomechanical coupling rate, $n_c$ is the mean number of photons in the optomechnical cavity, $\kappa$ is the decay rate of the optical cavity, and $\Gamma_m$ is the mechanical energy decay rate, $G$ is the light frequency shift per unit displacement, $x_{\text{zpf}}$ is the zero-point fluctuations, $\Omega_m = 2\pi\omega_m$ is the angular frequency of the mechanical resonator. From Eq. (4), it is evident that high finesse cavity, low effective mass and high mechanical quality of the mechanical resonator is critical to improve optomechanical cooperativity $C$.

The optomechanical cooperativity can be expressed as the single-photon cooperativity $C_0$ enhanced by the intra-cavity mean number of photons $n_c$

$$C = C_0 n_c, \quad C_0 = 4g_0^2 / \kappa \Gamma_m = 16\pi c x_{zpf}^2 F/(L\lambda^2 \Gamma_m), \quad (6)$$

where $c$ is the speed of light, $F$ and $L$ are the finesse and length of the cavity,

respectively, $\lambda$ is the wavelength of the laser. Note that optomechanical systems with large single-photon cooperativity might enable implementation of nonlinear optomechanical schemes for non-classical state generation such as photon antibunching [27]. By employing the presented phononically shielded multi-wavelength photonic-crystal membrane as the end mirror in a Fabry-Perot-type optomechanical system and using the simulated parameters of $x_{zpf} = 6.41 \times 10^{-16}$ m, $F = 1.53 \times 10^4$, $\lambda = 1566.3$ nm, $\Gamma_m = 2.16 \times 10^{-3}$ Hz, and a short cavity length of $L = 2$ μm, we can expect a large single-photon cooperativity of $C_0 \sim 1.97 \times 10^4$ at room temperature.

## 3. Potential applications

The presented phononically shielded multi-wavelength photonic-crystal membrane can interact with the optical cavity mode at different wavelengths via radiation pressure. This optomechanical interface can be employed to measure the optomechanical entanglement [28], and achieve the optical wavelength conversion of quantum states [29]. For both applications, a two-cavity optomechanical system required, as shown in Fig. 5. It is configured by two fixed end mirrors and a movable membrane mirror in the middle that also act as the mechanical resonator. The first mirror on the left and the membrane mirror forms the first Fabry-Perot cavity $C_1$, and the second mirror on the right combined with the membrane mirror forms a second Fabry-Perot cavity $C_2$.

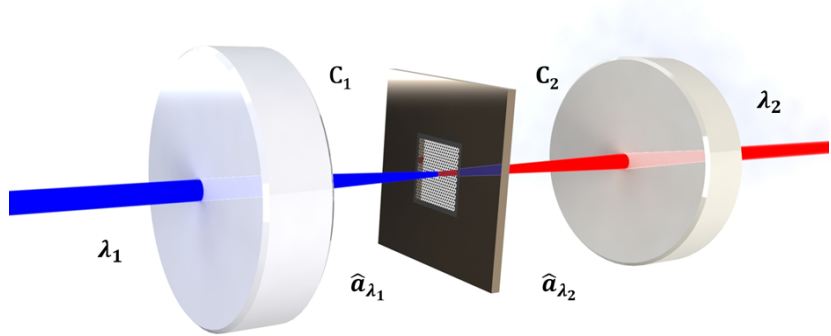

Fig. 5. Schematic diagram of the two-cavity optomechanics system. It consists of two fixed mirrors and a movable membrane in the middle. The first mirror on the left and the membrane constitutes the Fabry-Perot cavity C1, and the second mirror on the right with the membrane forms the second Fabry-Perot cavity C2.

To measure the optomechanical entanglement between the membrane resonator and the cavity field $\hat{a}_{\lambda_1}$ of $C_1$ at wavelength $\lambda_1$, the quantum state of the mechanical mode is mapped to an optical mode $\hat{a}_{\lambda_2}$ in $C_2$ via optomechanical interaction. In this case, both the position and the momentum of the membrane resonator can be measured by detecting the output field of $C_2$ at wavelength $\lambda_2$. Then, the quantum correlations between the two output fields of cavity $C_1$ and $C_2$ are measured to characterize the entanglement through the logarithmic negativity $E_N$ [30].

For the optical wavelength conversion of quantum states [29], the membrane resonator acts as an interface through which the quantum state is converted between different wavelengths of light fields. More precisely, we can map an optical field $\hat{a}_{\lambda_1}$ at wavelength $\lambda_1$ to a mechanical state, which in turn can be mapped to an optical field $\hat{a}_{\lambda_2}$ with wavelength $\lambda_2$.

## 4. Conclusions

We have proposed a phononically shielded multi-wavelength photonic-crystal SiN membrane for cavity quantum optomechanics. The membrane resonator features highly reflective at multi-wavelengths and high mechanical quality factor. The designed devices can be exploited to construct a compact optomechanical system with high optomechanical cooperativity. It may can find useful applications in cavity optomechanics to realize quantum frequency conversion, measurement of optomechanical entanglement, preparation of nonclassical optical and mechanical states and so on. In our future work, we will fabricate the designed membrane resonator and verify its optical and mechanical properties.

**Funding.** National Natural Science Foundation of China (11774209, 11804208, 12174232).

**Disclosures.** The authors declare no conflicts of interest.

**Data availability.** Data underlying the results presented in this paper are not publicly available at this time but maybe obtained from the authors upon reasonable request.

**References**

1. M. Aspelmeyer, T. J. Kippenberg, and F. Marquardt, eds., *Cavity Optomechanics: Nano- and Micromechanical Resonators Interacting with Light* (Springer Berlin Heidelberg, 2014).

2. W. P. Bowen, *Quantum Optomechanics*, 1st Edition (CRC Press, 2015).

3. S. Barzanjeh, A. Xuereb, S. Gröblacher, M. Paternostro, C. A. Regal, and E. M. Weig, "Optomechanics for quantum technologies," Nat. Phys. **18**, 15–24 (2022).

4. A. M. Jayich, J. C. Sankey, B. M. Zwickl, C. Yang, J. D. Thompson, S. M. Girvin, A. A. Clerk, F. Marquardt, and J. G. E. Harris, "Dispersive optomechanics: a membrane inside a cavity," New J. Phys. **10**, 095008 (2008).

5. J. D. Joannopoulos, J. N. Winn, and S. G. Johnson, *Photonic Crystals: Molding the Flow of Light - Second Edition* (Princeton University Press, 2011).

6. K. Sakoda, *Optical Properties of Photonic Crystals*, 2nd ed, Springer Series in Optical Sciences No. 80 (Springer, 2005).

7. V. Lousse, W. Suh, O. Kilic, S. Kim, O. Solgaard, and S. Fan, "Angular and polarization properties of a photonic crystal slab mirror," Opt. Express **12**, 1575 (2004).

8. S. Fan and J. D. Joannopoulos, "Analysis of guided resonances in photonic crystal slabs," Phys. Rev. B **65**, 235112 (2002).

9. C. H. Bui, J. Zheng, S. W. Hoch, L. Y. T. Lee, J. G. E. Harris, and C. Wei Wong, "High-reflectivity, high- $Q$

micromechanical membranes via guided resonances for enhanced optomechanical coupling," Appl. Phys. Lett. **100**, 021110 (2012).

10. S. Kini Manjeshwar, K. Elkhouly, J. M. Fitzgerald, M. Ekman, Y. Zhang, F. Zhang, S. M. Wang, P. Tassin, and W. Wieczorek, "Suspended photonic crystal membranes in AlGaAs heterostructures for integrated multi-element optomechanics," Appl. Phys. Lett. **116**, 264001 (2020).

11. F. Zhou, Y. Bao, J. J. Gorman, and J. R. Lawall, "Cavity Optomechanical Bistability with an Ultrahigh Reflectivity Photonic Crystal Membrane," Laser & Photonics Rev. **17**, 2300008 (2023).

12. R. A. Norte, J. P. Moura, and S. Gröblacher, "Mechanical Resonators for Quantum Optomechanics Experiments at Room Temperature," Phys. Rev. Lett. **116**, 147202 (2016).

13. T. Antoni, A. G. Kuhn, T. Briant, P.-F. Cohadon, A. Heidmann, R. Braive, A. Beveratos, I. Abram, L. L. Gratiet, I. Sagnes, and I. Robert-Philip, "Deformable two-dimensional photonic crystal slab for cavity optomechanics," Opt. Lett. **36**, 3434 (2011).

14. A. R. Agrawal, J. P. Manley, D. Allepuz-Requena, and D. J. Wilson, "A suspended focusing Si3N4 metamirror for integrated cavity optomechanics," in *Frontiers in Optics + Laser Science 2023 (FiO, LS) (2023), Paper FTh3B.4* (Optica Publishing Group, 2023), p. FTh3B.4.

15. J. P. Moura, R. A. Norte, J. Guo, C. Schäfermeier, and S. Gröblacher, "Centimeter-scale suspended photonic crystal mirrors," Opt. Express **26**, 1895–1909 (2018).

16. G. Enzian, Z. Wang, A. Simonsen, J. Mathiassen, T. Vibel, Y. Tsaturyan, A. Tagantsev, A. Schliesser, and E. S. Polzik, "Phononically shielded photonic-crystal mirror membranes for cavity quantum optomechanics," Opt. Express **31**, 13040 (2023).

17. A. R. Agrawal, J. Manley, D. Allepuz-Requena, and D. J. Wilson, "Focusing membrane metamirrors for integrated cavity optomechanics," Optica **11**, 1235 (2024).

18. F. Zhou, Y. Bao, J. J. Gorman, and J. R. Lawall, "Ultrahigh reflectivity photonic crystal membranes with optimal geometry," APL Photonics **9**, 076120 (2024).

19. Y. Tsaturyan, A. Barg, E. S. Polzik, and A. Schliesser, "Ultracoherent nanomechanical resonators via soft clamping and dissipation dilution," Nature Nanotech **12**, 776–783 (2017).

20. Z. Wang, E. S. Polzik, and D. G. Enzian, "Characterization of high-Q ultralight membranes towards novel optomechanical designs," University of Copenhagen (2022).

21. D. F. Santos, A. Guerreiro, and J. M. Baptista, "Numerical investigation of a refractive index SPR D-type optical fiber sensor using COMSOL multiphysics," Photonic Sens **3**, 61–66 (2013).

22. H. R. Philipp, "Optical Properties of Silicon Nitride," J. Electrochem. Soc. **120**, 295 (1973).

23. Y. Tsaturyan, A. Barg, A. Simonsen, L. G. Villanueva, S. Schmid, A. Schliesser, and E. S. Polzik, "Demonstration of suppressed phonon tunneling losses in phononic bandgap shielded membrane resonators for high-Q optomechanics," Opt. Express **22**, 6810–6821 (2014).

24. C. Reetz, R. Fischer, G. G. T. Assumpção, D. P. McNally, P. S. Burns, J. C. Sankey, and C. A. Regal, "Analysis of Membrane Phononic Crystals with Wide Band Gaps and Low-Mass Defects," Phys. Rev. Appl. **12**, 044027 (2019).

25. D. S. Bindel and S. Govindjee, "Elastic PMLs for resonator anchor loss simulation," International Journal for Numerical Methods in Engineering **64**, 789–818 (2005).

26. A. Frangi, A. Bugada, M. Martello, and P. T. Savadkoohi, "Validation of PML-based models for the evaluation of anchor dissipation in MEMS resonators," European Journal of Mechanics - A/Solids **37**, 256–265 (2013).

27. K. Børkje, "Critical quantum fluctuations and photon antibunching in optomechanical systems with large single-photon cooperativity," Phys. Rev. A **101**, 053833 (2020).

28. J. Chen, M. Rossi, D. Mason, and A. Schliesser, "Entanglement of propagating optical modes via a mechanical interface," Nat Commun **11**, 943 (2020).

29. L. Tian and H. Wang, "Optical wavelength conversion of quantum states with optomechanics," Phys. Rev. A **82**, 053806 (2010).

30. D. Vitali, S. Gigan, A. Ferreira, H. R. Böhm, P. Tombesi, A. Guerreiro, V. Vedral, A. Zeilinger, and M. Aspelmeyer, "Optomechanical Entanglement between a Movable Mirror and a Cavity Field," Phys. Rev. Lett. **98**, 030405 (2007).